\documentclass[preprint,12pt,authoryear]{elsarticle}

\usepackage{amssymb}
\usepackage{amsmath}
\usepackage[mathlines]{lineno}

\journal{Advances in Water Resources}

\begin{document}
\begin{frontmatter}
\title{Impact of porous media heterogeneity on convective mixing in a Rayleigh-B{\'e}nard instability}
\author[authorIDAEA]{Rima Benhammadi}\ead{rima.benhammadi@idaea.csic.es}
\author[authorIDAEA]{Juan J. Hidalgo}\ead{juanj.hidalgo@idaea.csic.es}
\cortext[cor1]{Corresponding author}
\address[authorIDAEA]{Institute for Environmental Assessment and Water
  Research, Spanish National Research Council, Barcelona, Spain}
%
%
\begin{abstract} 
This work studies the effect of the heterogeneity of a porous medium on convective mixing. We consider a system in which a Rayleigh-Bénard instability is triggered by a temperature difference between the top and bottom boundaries. Heterogeneity is represented by multi-Gaussian log-normally distributed permeability fields. We explore the effect of the Rayleigh number, the variance and correlation length of the log-permeability field on the fingering patterns, heat flux, mixing state and flow structure. Heat flux increases for all heterogeneous cases compared to the homogeneous ones. When heterogeneity is weak and the horizontal correlation length small, flux exhibits minimal sensitivity to the variance of the log-permeability. When the correlation length increases, flux increases proportionally to the log-permeability variance.

The mixing state is evaluated through the temperature variance and the intensity of segregation. Both take higher values, compared to their homogeneous analogues, when the correlation length and the variance of the permeability are increased. This indicates that even if heat flux increases, the system is less well mixed.

The flow structure shows that in homogeneous and weakly heterogeneous cases there is a relation between the location of high strain rates and stagnation points, while for strongly heterogeneous cases, high strain rate zones are linked to high permeability areas near the boundaries, where temperature plumes originate. The interface width tends to decrease as the variance and the correlation length of the permeability field are augmented, suggesting that the interface undergoes greater stretching in heterogeneous porous media. 
\end{abstract}
%
%
%
%
\begin{highlights}
\item Heterogeneity in the permeability field increases heat flux
\item Heterogeneity induces fluid segregation 
\item Formation of thermal plumes linked to high permeability areas
\end{highlights}
%
%
%
%
\end{frontmatter}
%
%
%
%
\section{Introduction}
Convective mixing is a critical process in a wide range of natural and engineered systems, influencing phenomena such as seawater intrusion in coastal aquifers, leachate infiltration in waste disposal sites, salt transport in agricultural practices, groundwater movement in salt formations for radioactive waste storage, geothermal energy extraction, and carbon capture and storage (CCS) \citep{Simmons2001}. Recent advancements in this field, however, indicate the complexity of convective dynamics and the need to take other factors into account such as permeability heterogeneity and anisotropy \citep{article_Green, chen2013}. At its core, convective mixing arises from density gradients between two miscible fluids, where the destabilising effect of the gradient leads to convective flows. These flows significantly accelerate mass and energy transfer, outperforming diffusion-based processes \citep{ching_convective_2017}. The resulting convective structures, such as vortices and stagnation points, enhance mixing through mechanisms like stretching and compression of fluid interfaces \citep{hidalgo_scaling_2012, hidalgo_dissolution_2015}.

Much of the existing research on convective mixing has focused on homogeneous media, where material properties such as permeability and porosity are assumed to be uniform. In such systems, convection typically manifests through well-defined patterns, with the instability driving the formation of convection rolls, fingers, and plumes. However, real-world systems are rarely homogeneous, and understanding how heterogeneity, whether in permeability, porosity, or other material properties, affects convective dynamics is still an open challenge. While the dynamics of convective mixing in homogeneous media are well-established, the influence of heterogeneity on these processes is less understood and requires deeper examination.

Heterogeneity in porous media is known to profoundly affect convective instability, altering the onset of convection, the flow dynamics, and the efficiency of heat and mass transport. Numerical studies have shown that variations in permeability can reduce the onset time for convection, especially in high-permeability regions, and lead to complex flow regimes such as fingering, channeling, and dispersive flows, all of which depend on the statistical properties of permeability, such as correlation length and variance \citep{article_Farajzadeh, ranganathan2012, article_Green, Wang2025}. These effects can change the interaction between convective structures, leading to variations in finger width, flow patterns, and the extent of mixing. Importantly, while the role of permeability is well-studied, the impact of other factors, such as the anisotropy of the medium or the presence of geometric constraints, on convective behaviour remains underexplored \citep{Li2019}. Despite the recognised significance of heterogeneity in natural systems, research directly addressing how it impacts Rayleigh-Bénard convection in porous media is still limited.

Studies examining the effects of heterogeneity in specific applications, such as CO$_2$ sequestration, have provided valuable insights into how variations in permeability can influence processes like CO$_2$ trapping and dissolution fluxes. For example, anisotropic sedimentary rocks have been found to exhibit enhanced CO$_2$ dissolution due to better connectivity of permeability structures \citep{DePaoli2016, Ershadnia2021, Zhang2024}. However, the influence of permeability correlation length on fluxes remains poorly investigated with the exception of \citet{ranganathan2012} and \citet{Wang2025} who numerically studied that in the context of CO$_2$ storage. In spite of the growing body of research in these applications, the majority of studies on density-driven flows have focused on Rayleigh-Taylor instabilities, with relatively few addressing Rayleigh-Bénard instability, especially in heterogeneous porous media. For instance, numerical studies of the Horton-Rogers-Lapwood (HRL) problem \citep{1945JAP....16..367H, 1948PCPS...44..508L}, a Rayleigh-Bénard-like convection problem in porous media, have revealed that scalar dissipation rates and the efficiency of mixing in homogeneous porous media are sensitive to interfacial processes \citep{hidalgo_mixing_2018}. Meanwhile, experimental and numerical investigations have provided important insights into convection in confined geometries. \citet{Letelier2019} used numerical simulations in Hele-Shaw cells to show how geometric constraints such as sidewall conduction and finite aspect ratio alter classical Rayleigh-Bénard scaling, reducing heat transport efficiency at high Rayleigh numbers. \citet{Ulloa2025} further extended this work by analysing energy transfer and mixing efficiency in thermally driven flows, identifying how geometric confinement influences boundary layer development, protoplume formation, and the organisation of convective structures. These studies, along with recent imaging-based experiments such as those by \cite{Sin2024, Benhammadi2025}, highlight the critical role of geometry in shaping the dynamics and transport properties of convection in porous media.

One of the key challenges in the study of convective mixing in heterogeneous media is understanding how the geometric and material heterogeneities alter classical scaling laws and mixing behaviours. Nevertheless, a critical gap remains in our understanding of how heterogeneity in porous media affects the scaling laws and convective dynamics of Rayleigh-Bénard convection. 

In this paper, we hence propose to address this gap by investigating convective mixing in heterogeneous porous media within the context of the HRL problem. This problem models heat transport driven by Rayleigh-Bénard instability, where convection is triggered by a temperature difference between the top and bottom boundaries of a porous medium. We compare simulations in homogeneous and heterogeneous media, focusing on key metrics such as fluxes, the mixing state and the segregation intensity.

The structure of the manuscript is as follows: Section 2 describes the methodology, Section 3 presents the simulation results and corresponding discussions, and Section 4 concludes with a summary and final remarks.
%
%
\section{Methodology}
We numerically study convective mixing in the case of the HRL problem in heterogeneous porous media. The HRL problem examines the variable-density heat transport and the convective patterns which arise due to a Rayleigh-Bénard instability caused by the density differences between the top and bottom boundaries where the scalar temperature is constant.
%
%
\subsection{Governing equations}
Under the assumptions of incompressible fluid and the Boussinesq approximation, the dimensionless governing equations for variable-density single-phase flow in a porous medium are \citep{riaz_2006}
\begin{align} 
\label{eq:continuity}
  \nabla \cdot \mathbf{q} = 0 
\end{align} 
\begin{align} 
\label{eq:Darcy}
\mathbf{q} = - k \left(\nabla p - \rho \mathbf{\hat{e}}_g\right) 
\end{align} 
\begin{align} 
\label{eq:RD}
 \frac{\partial T}{\partial t} + \mathbf{q} \cdot \nabla T = \frac{1}{Ra} \nabla^2 T,
\end{align} 
where $\mathbf{q}$ is  Darcy velocity, $p$ a scaled pressure referred to a hydrostatic datum, $\rho$ is the fluid's density and a function of $T$, temperature, $\mathbf{\hat{e}}_g$ is a unit vector pointing in the direction of gravity, $k$ is the medium's permeability, $Ra$ the Rayleigh number, and $t$ is time. The Rayleigh number is defined as
\begin{align} 
\label{eq:Ra}
Ra = \frac{q_b H}{\phi D_m},
\end{align} 
with $H$ a characteristic length taken as the domain height, $\phi$ porosity, $D_m$ the thermal conductivity coefficient and $q_b = k_{g}\Delta \rho g/\mu$ a buoyancy velocity, where $\mu$ is the fluid's viscosity, $g$ the gravity acceleration, $\Delta \rho$ the maximum density difference in the system and $k_{g}$ the geometric mean of the permeability field. We assume a linear dependence of density with $T$ so that $\rho = -T$ in dimensionless form.

This problem is solved within a rectangular $2 \times 1$ domain. All boundaries are no flow, zero temperature gradient ($\mathbf{q} \cdot \mathbf{n} = 0$; $\nabla T \cdot \mathbf{n} = 0$) except the top and bottom boundaries where $T=0$, $T=1$ is prescribed respectively. The maximum density difference is, therefore, given by $\Delta \rho = \rho(T=0) - \rho(T=1)$.

The porous media heterogeneity is represented by a multi-Gaussian log-normally distributed random field with zero mean and variance $\sigma_{\log k}^2$, with a Gaussian variogram
\begin{align}
  \label{eq:variogram}
  \text{Var}(x) = \sigma_{\log k}^2 \left[ 1 - e^{-\left(h_{x}/\lambda_{x}\right)^{2} - \left(h_{z}/\lambda_{z}\right)^{2}} \right],
\end{align}
where $h_{i}$, $\lambda_{i}$ is the separation and correlation length in direction $i$ with $i=x, z$.

A parametric study is conducted considering different values for the Rayleigh number $(1e2 \leq Ra \leq 1e4)$, the permeability field variance $(\sigma_{\log k}^2 = 0.25, 1, 2, 3)$ and correlation length. We take $\lambda_{x} = 2 \lambda_{z} = 0.02, 0.2, 0.7$ and $\lambda_{x} = 7\lambda_{z} = 0.7$. For each case, five realisations are performed. Additionally, the homogeneous cases ($\sigma_{\log k}^{2} = 0$)  are simulated to provide a comparative basis for the heterogeneous cases.

Numerical simulations were performed using the variable-density flow and transport solver \texttt{rhoDarcyFoam} open-source computational framework SECUReFoam \citep{Icardi2023}, based on the finite-volume library OpenFOAM \citep{Weller1998}. The details of the numerical simulations are described in appendix~\ref{app:numerics}.
%
%
\section{Results and discussion}
In this section, we explore convective mixing in the HRL problem in heterogeneous media from both qualitative and quantitative perspectives. Qualitatively, we examine the temperature field under varying correlation lengths and variances of the permeability field $\sigma_{\log k}^2$. Quantitatively, we analyse heat fluxes as a function of the Rayleigh number $Ra$ and $\sigma_{\log k}^2$, investigate the fixed interface dynamics through the interface width, and assess the mixing state by examining the temperature probability density functions and the intensity of segregation.
%
%
\subsection{Instability patterns} 
\subsubsection{Effect of Ra and $\lambda_{x}$}
We analyse the influence of heterogeneity on temperature patterns, first by investigating the impact of convective instability through $Ra$, and then by examining the role of heterogeneity at a fixed $Ra$.

\begin{figure}
    \centering   
\includegraphics[width=\linewidth]{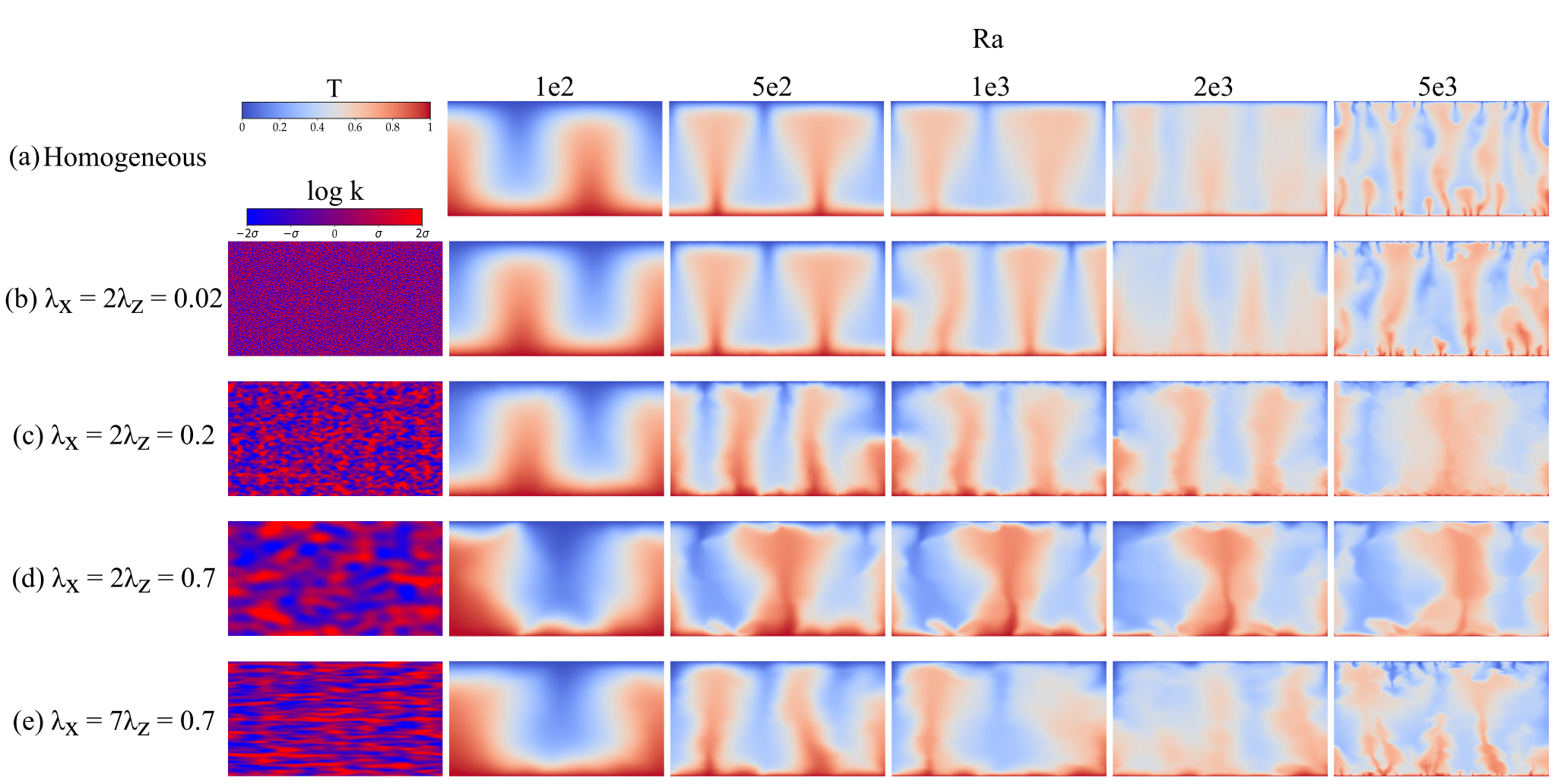}
\caption{Log-permeability field (leftmost column) and time-averaged temperature maps at convective regime ($60<t<250$) for the homogeneous (a) and $\sigma_{\log k}^2=1$ cases (b--e) with varying $Ra, \lambda_{x}$, and $\lambda_{z}$. Increasing $Ra$ increases the number of patterns in the homogeneous (a) and heterogeneous case (b) thus increasing the interfacial area between the two fluids in contact, unlike in the other heterogeneous cases where merging occurs faster and fingers become distorted and dispersive (particularly in cases (c) and (d)). Instantaneous temperature snapshots are shown in figures 1--4 of the supplementary material for times $t=20,60,100,140$}
\label{fig:CvsRa}
\end{figure}

The homogeneous simulations (Figure~\ref{fig:CvsRa}a) reveal that for $Ra < 2000$, the instability patterns extend throughout the entire domain, forming stable convection cells in agreement with previous observations in squared \citep{Graham1994} and rectangular \citep{hidalgo_mixing_2018} domains. As $Ra$ increases, the system transitions into a chaotic convection regime characterised by columnar patterns \citep{Hewitt2013}. In the heterogeneous cases (Figure~\ref{fig:CvsRa}b -- e), the patterns become more dispersive. For a given $\sigma^2_{\log k}$, the higher the $Ra$, the stronger the dispersion. For low $Ra$, when the instability is organised into convective cells, heterogeneity distorts the shape of the cells (especially for high $\lambda_{x}$) but the overall shape of the pattern remains unaltered. As $Ra$ increases, the characteristic size of the pattern decreases and becomes more similar to the heterogeneity structure making the interaction between the instability and heterogeneity stronger. This interaction modifies the temperature pattern reducing the number of convection cells or columns with respect to the homogeneous case, although the trend is not always monotonic. At intermediate values of $\lambda_{x}$, an increase in the convection cells is observed for $Ra=5e2, 1e3$.

From Figure \ref{fig:CvsRa}a and b, it is apparent that whether the permeability is homogeneous or strongly heterogeneous, the number of fingers increases with $Ra$, with the interface becoming more stretched at higher $Ra$. However, for larger $\lambda_{x}$ (Figure \ref{fig:CvsRa}c to e), the merging of fingers occurs earlier as $Ra$ increases. Interestingly, the merging is less rapid when the anisotropy ratio ($\lambda_{x}/\lambda_{z}$) is big (compare Figure \ref{fig:CvsRa}d to Figure \ref{fig:CvsRa}e), suggesting that small anisotropy ratios accelerate the merging process. For small anisotropy ratio (Figure \ref{fig:CvsRa}d), merging occurs even at lower $Ra$, and the fingers assume a mushroom-like shape.

\subsubsection{Effect of $\sigma^2_{\log k}$}

\begin{figure}
\centering
\includegraphics[width=\linewidth]{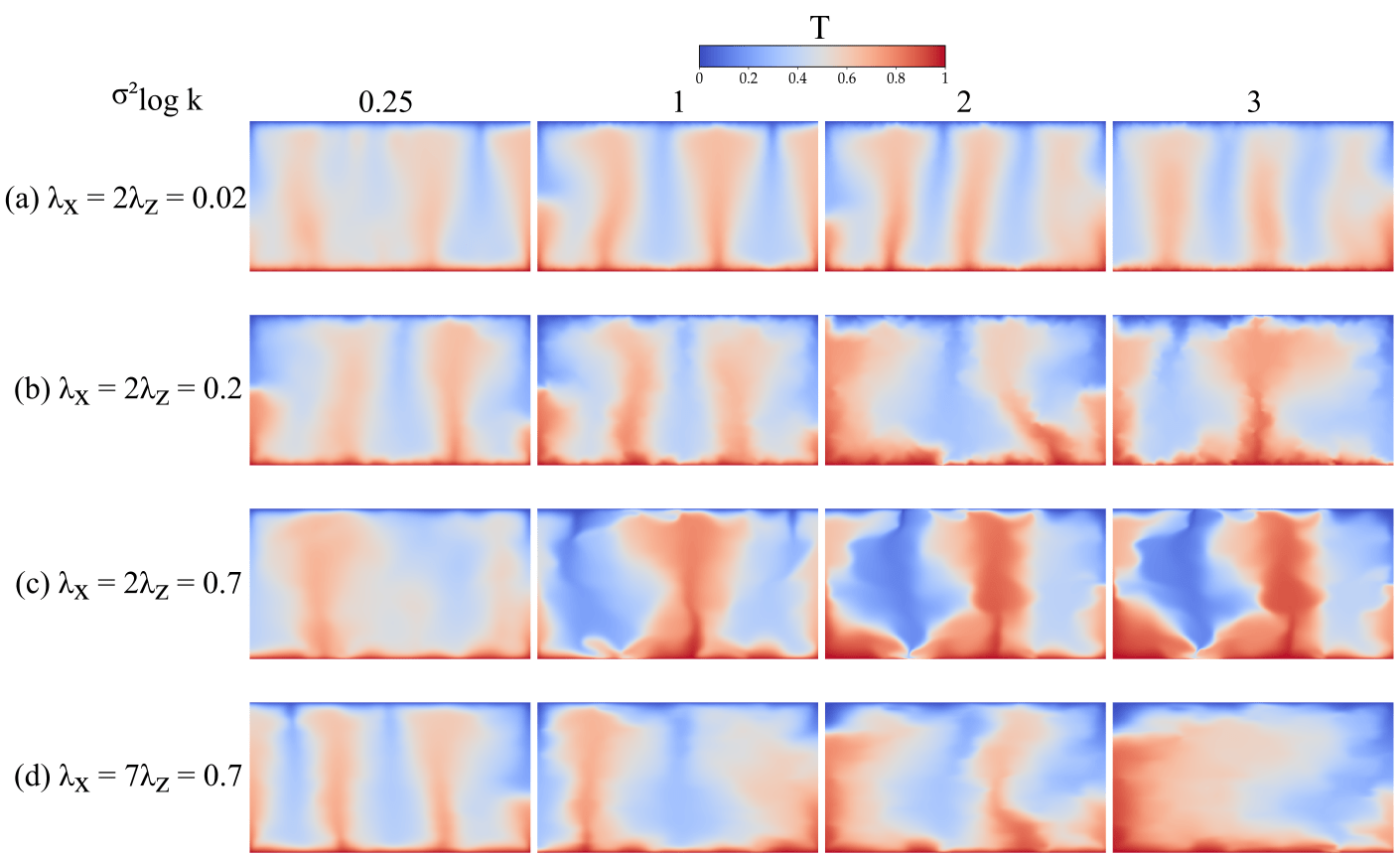}
\caption{Time-averaged temperature maps at convective regime ($60<t<250$)  for $Ra=1e3$, different $\sigma_{\log k}^{2}, \lambda_{x}$ and $\lambda_{z}$. Increasing $\sigma_{\log k}^{2}$ and $\lambda_x$ (especially with a smaller anisotropy ratio), promotes faster merging of the fingers. Instantaneous temperature snapshots are shown in the supplementary material for times $t=20,60,100,140$}
\label{fig:CvsL}
\end{figure}

Figure \ref{fig:CvsL} illustrates the temperature patterns for $Ra = 1e3$ and varying $\lambda_{x}, \lambda_{z}$ and $\sigma_{\log k}^2$.  It is observed that as $\sigma_{\log k}^2$ increases, the number of columns decreases. The reduction in the number of columns is faster for longer $\lambda_{x}$, which means that the barrier effect of the heterogeneity induces a faster merging of the plumes. As seen earlier in Figure~\ref{fig:CvsRa}, the patterns also become more dispersive, except for small $\lambda_{x}$ (Figure~\ref{fig:CvsRa}b) for which the dispersive effect is weaker. This behaviour aligns with the findings of \cite{ranganathan2012} for a Rayleigh-Taylor instability, who noted that increasing $\sigma_{\log k}^2$ with a small $\lambda_{x}$ induces flow dispersion, whereas increasing $\sigma_{\log k}^2$ with a large $\lambda_{x}$ promotes channeling. For a given, sufficiently large $\lambda_{x}$ (Figure~\ref{fig:CvsRa}d, e), a smaller anisotropy ratio has a less distorting effect. Thin horizontal barriers (Figure~\ref{fig:CvsRa}e) modify the pattern more than thick short ones (Figure~\ref{fig:CvsRa} d) for all the $\sigma_{\log k}^{2}$ values considered.
%

%
%
\subsection{Mixing and heat flux}
Mixing is usually assessed by the evolution of the variance of the temperature $\sigma_{T}^{2}$ \citep{Kapoor1998}. For the HRL problem, the evolution of $\sigma_{T}^{2}$ is given by
\begin{align} 
\label{eq:fluxvart}
\frac{\partial \sigma_{T}^{2}}{\partial t} = \frac{\partial (\langle T^{2}\rangle - \langle T \rangle^{2}) }{\partial t} = 2\left(T_{\text{top}}F_{\text{top}} - T_{\text{bottom}}F_{\text{bottom}}\right) - 2\langle T\rangle(F_{\text{top}} - F_{\text{bottom}})- 2\langle{\chi}\rangle,
\end{align}
where $\langle \cdot \rangle$ indicates spatial integration, $T_{\text{top}}, T_{\text{bottom}}$ are the temperatures prescribed at the top and bottom boundaries, $F_{\text{top}}$, $F_{\text{bottom}}$ the flux through the top and bottom boundaries and
\begin{align}
  \label{eq:chi}
  \langle \chi \rangle = \frac{1}{Ra}\int_{\Omega} \left|\nabla T \right|^{2} d \Omega
\end{align}
is the total scalar dissipation rate. Replacing $T_{\text{bottom}} = 1$ and $T_{\text{top}} = 0$ in \eqref{eq:fluxvart} and using that at the steady state, $F_{\text{top}} = -F_{\text{bottom}}$, we obtain
\begin{align} 
\label{eq:flux}
 \langle \chi \rangle =  F_{\text{bottom}}.
\end{align} 

Therefore, the scalar dissipation rate gives information about the flux through the prescribed temperature boundaries once the system arrives at a steady state. Except for low $Ra$, for which patterns reach a stable configuration, the system does not attain a steady state. The total flux and $\langle \chi \rangle$ oscillate around a mean value after convection is fully developed. Therefore, they are quantified as the temporal mean in the convection-dominated regime. To account for the heterogeneity variablity, five realisations of each case were performed and the ensemble average calculated.

\begin{figure}
    \centering    
    \includegraphics[width=0.5\linewidth]{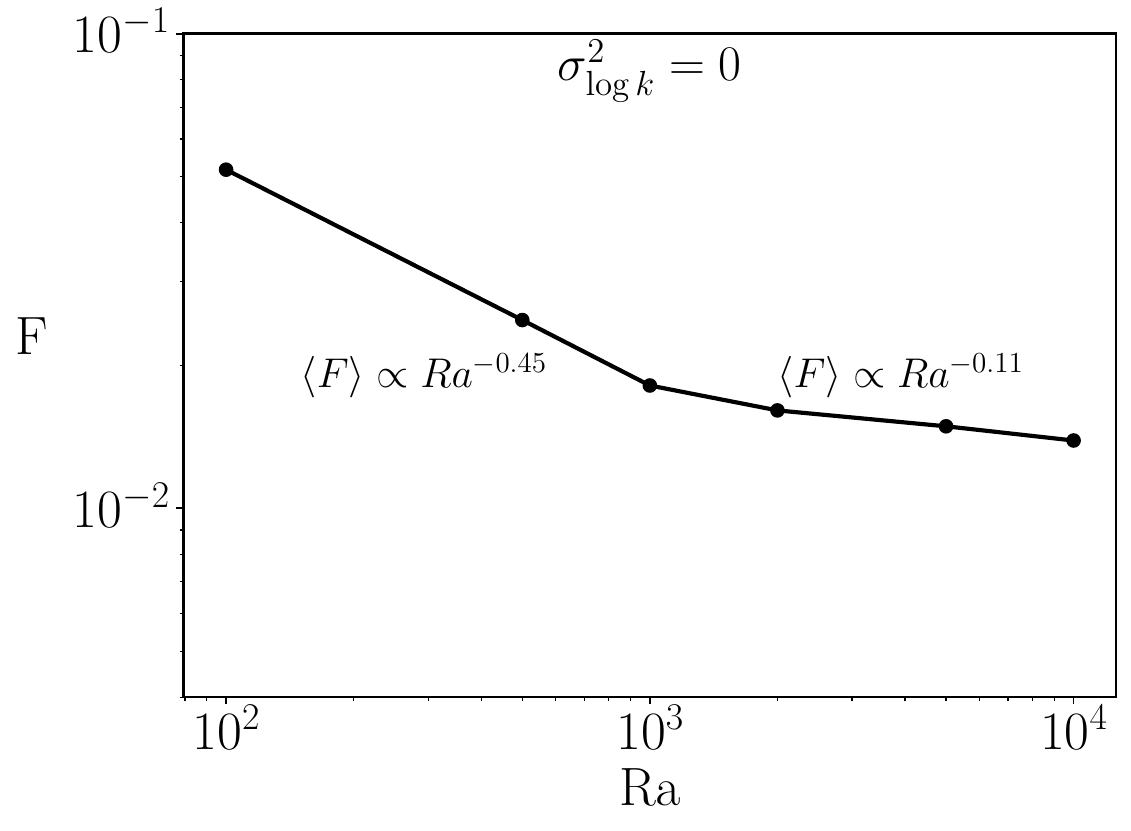}   
    \caption{Heat flux through the boundary versus $Ra$ for the homogeneous case. A transition in the scaling is observed for $Ra> 10^{3}$ corresponding to the transition from organised patterns to chaotic patterns. Figures of heat fluxes versus time are shown in figures 5 and 6 of the supplementary material} \label{fig:FluxRahom}
\end{figure}

\begin{figure}
    \centering    
    \includegraphics[width=\linewidth]{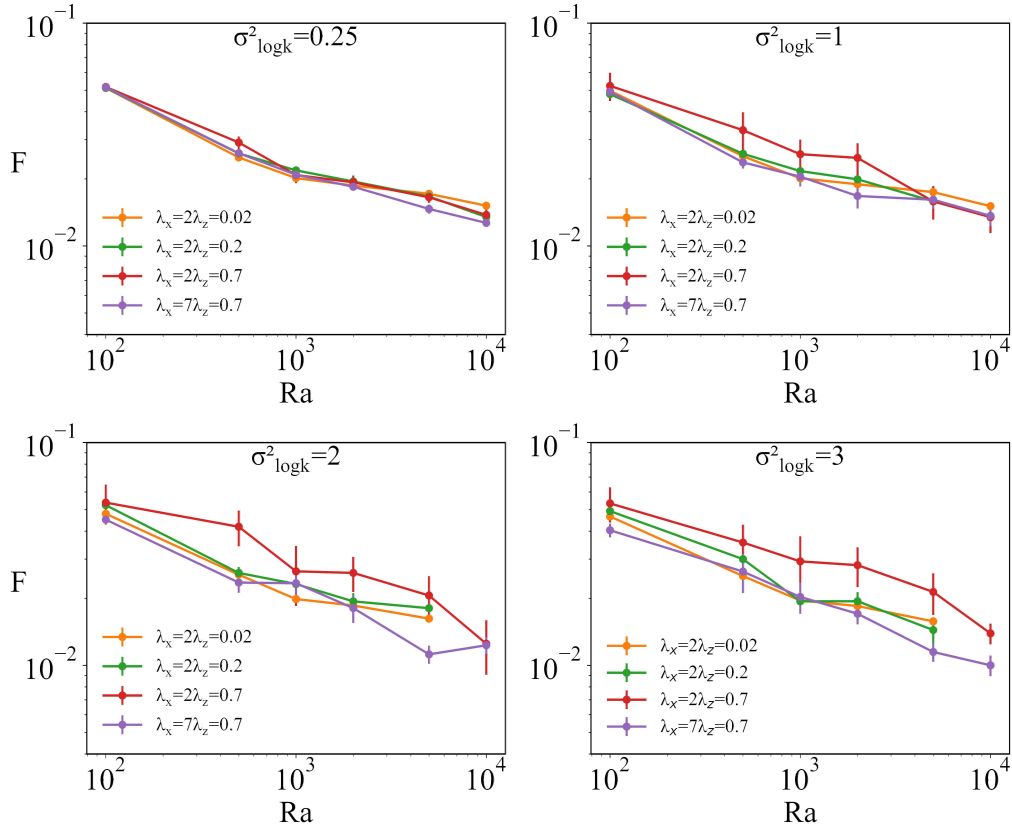}   
    \caption{Flux versus $Ra$ for different correlation lengths and different variances of the permeability field. Flux is computed as the average of five realisations. Error bars show the standard deviation of the ensemble average. A higher $\sigma^2_{\log k}$ increases the error and causes greater separation between cases with different $\lambda_x$. The fluxes are also sensitive to $\lambda_x$: the larger it is, the higher the fluxes. For the same $\lambda_x$, cases with a smaller anisotropy ratio have bigger fluxes than those with a larger anisotropy ratio. Figures of heat flux versus time fro the heterogeneous cases are shown in figures 5 and 6 of the supplementary material}
   \label{fig:FluxRa}
\end{figure}

Figures \ref{fig:FluxRahom} and \ref{fig:FluxRa} show, respectively, the behaviour of the flux for the homogeneous case and ensemble average of the total flux for the heterogenous cases (equivalent to $\langle \chi \rangle$). The difference between the homogeneous and heterogenous cases is small for weakly heterogeneous cases $(\sigma_{\log k}^2= 0.25, 1)$. However, as $\sigma_{\log k}^2$ increases, the flux associated with the highest $\lambda_{x}$ and the smallest anisotropy ratio becomes noticeably higher. A comparison between the two cases with $\lambda_{x}=0.7$ reveals that a lower anisotropy ratio corresponds to a higher flux. Furthermore, Figure \ref{fig:FluxRa} shows that as $\lambda_{x}$ increases, the fluxes rise more sharply with $\sigma_{\log k}^2$. Additionally, flux decreases as $Ra$ increases. This suggests that stronger heterogeneity, represented by smaller $\lambda_{x}$ values and higher $\sigma_{\log k}^2$, reduces heat fluxes when compared to cases with larger $\lambda_{x}$ values. These results align with the findings of \cite{article_Farajzadeh} and \cite{ranganathan2012} for one-sided instabilities in the context of CO$_{2}$ dissolution.

Regarding the flux scaling with $Ra$, the homogeneous case and the weakly heterogeneous case with small $\lambda_{x}$ present a transition to a convection-dominated regime around $Ra=1000$ in which flux goes from scaling as $Ra^{-0.45}$ to $Ra^{-0.11}$ for the homogeneous case (see figure \ref{fig:FluxRahom}) and, for instance, from $Ra^{-0.35}$ to $Ra^{-0.13}$ for $\lambda_x=2\lambda_z=0.02$ (orange line in Figure \ref{fig:FluxRa}). As heterogeneity increases, the convection-dominated regime is not observed and flux always depends on $Ra$. Exponents range between $-0.38$ and $-0.27$ with little dependence on the anisotropy ratio or correlation length.

%
%
\subsection{Interface compression and velocity structure}
\begin{figure}
  \centering
  \includegraphics[width=\linewidth]{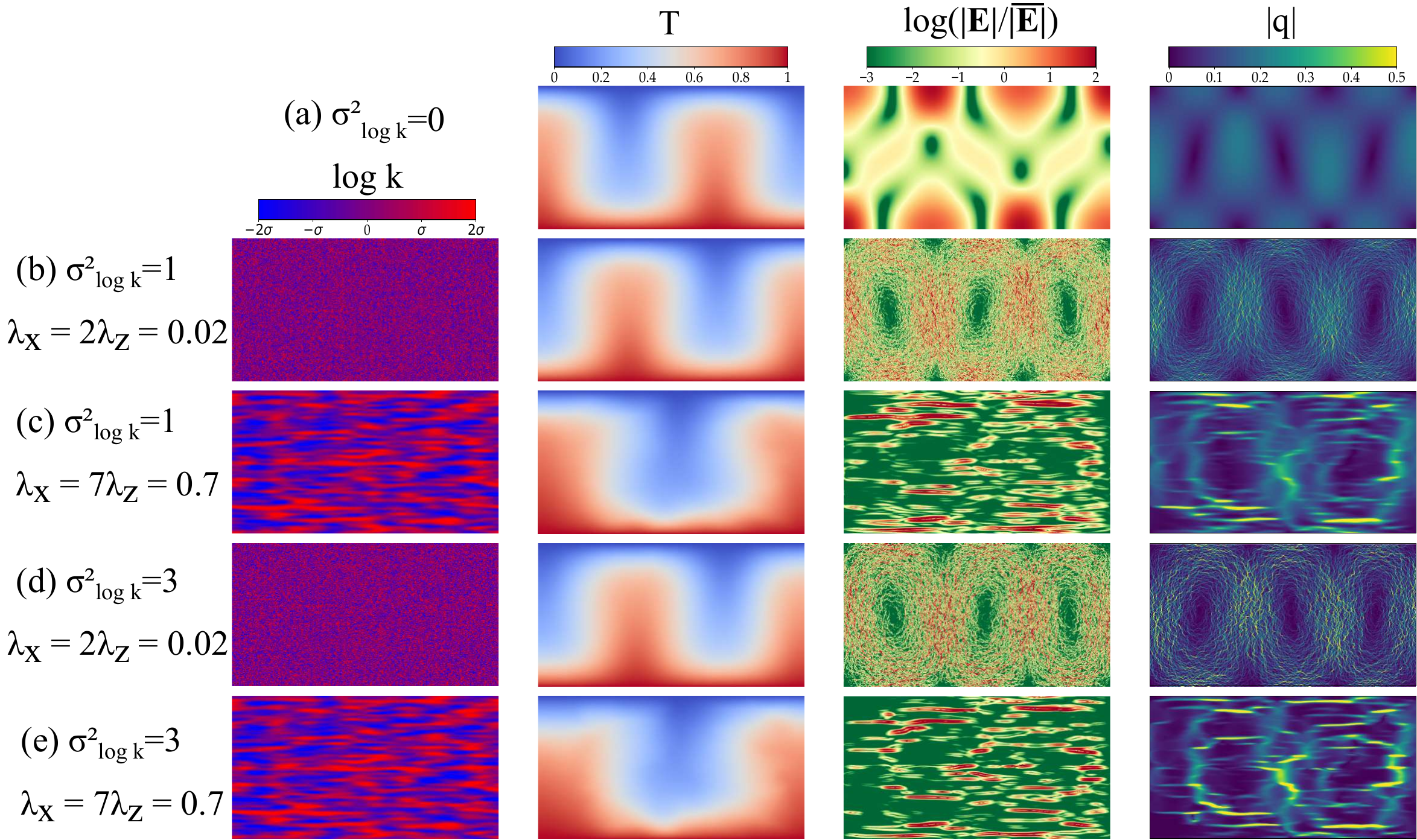}
  \caption{Log-permeability ($\log k$), temperature $T$ and logarithm of the determinant of the strain tensor $\mathbf{E}$ normalised by its mean for $Ra=1e2$, $\sigma_{\log k}^{2}=1,3$ and different $\lambda_{x}$, $\lambda_{z}$. Values are time averaged in the convective regime ($60 < t <250$)}
  \label{fig:kTe100}
\end{figure}

\begin{figure}
  \centering
  \includegraphics[width=\linewidth]{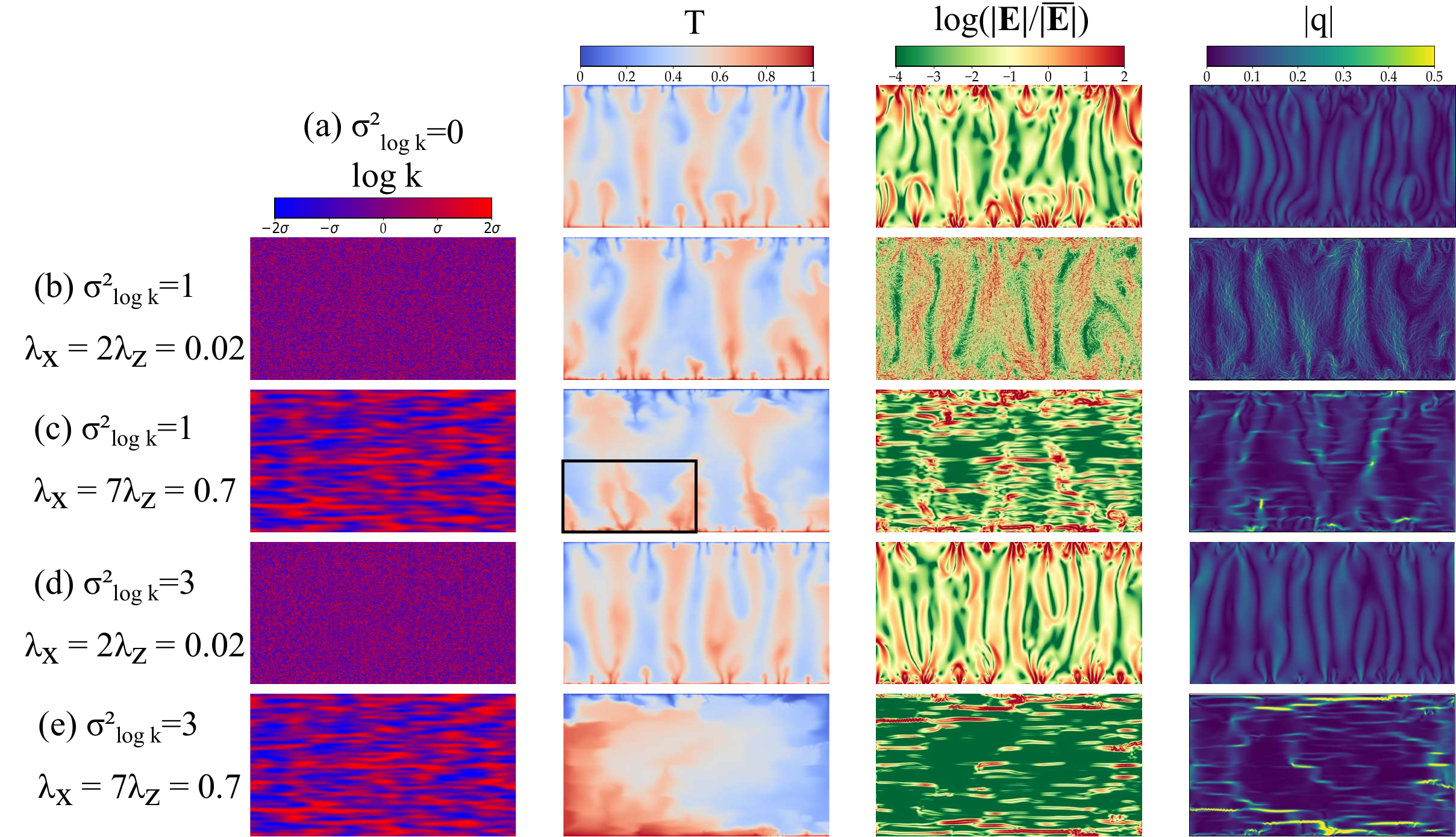} 

    \caption{Log-permeability ($\log k$), temperature $T$ and logarithm of the determinant of the strain tensor normalised by its mean for $Ra=5e3$, $\sigma_{\log k}^{2}=1,3$ and different $\lambda_{x}$, $\lambda_{z}$. Values are time averaged in the convective regime ($60 < t <250$). The rectangle in the temperature plot of case (c) indicates the location of the zoomed-in region shown in figure \ref{fig:strain_detail}}
    \label{fig:kTe5000}
\end{figure}

\begin{figure}
  \centering  
  \includegraphics[width=\linewidth]{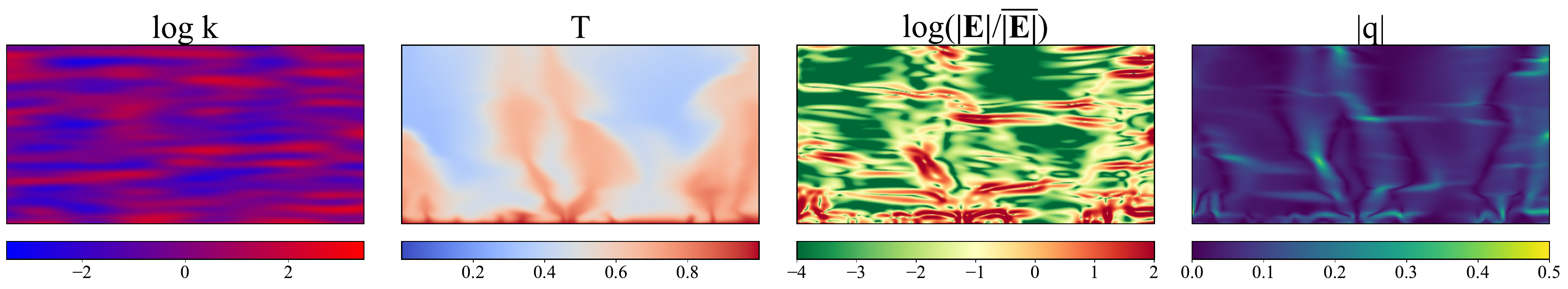}   
    \caption{Detail of the log-permeability, temperature, logarithm of the determinant of the strain tensor $\mathbf{E}$ normalised by its mean and velocity at the bottom half left part of the cell for the case with $Ra=5e3$, $\sigma^2_{\log k}=1$ and $\lambda_x=7\lambda_z=0.7$ (see Figure~\ref{fig:kTe5000}c for the location of the zoomed-in area). Plumes develop at stagnation points where high strain is found. The location of the stagnation points coincides with high permeability regions}
\label{fig:strain_detail}
\end{figure}

In convection-dominated system, there exists a relation between the flux, scalar dissipation rate and the behaviour of the fluid interface near flow stagnation points \citep{hidalgo_dissolution_2015}. In the HRL problem, this fluid interface is located at the top and bottom boundaries where temperature is prescribed. The thickness of the fluid interface $s$ is given by \citep{Villermaux_2012, Le_Borgne_2013}
\begin{align}
  \label{eq:s}
  \frac{1}{s} \frac{\partial s}{\partial t} = -\gamma + \frac{1}{Ra} \frac{1}{s^{2}},
\end{align}
where $\gamma$ is the dimensionless compression rate, which is given by the symmetric part of the strain tensor \citep{Ottino1989}
\begin{align}
  \label{eq:strain_tensor}
  \mathbf{E} = \frac{1}{2}\left(\nabla \mathbf{q} + \nabla \mathbf{q}^{T} \right) = \left[\begin{array}{cc} \gamma & 0 \\ 0 & -\gamma \end{array} \right]
\end{align}
At the steady state, compression and diffusion equilibrate and the interface thickness has a length $s_{B}$, known as the Batchelor scale \citep{Batchelor_1959, Villermaux_2006},
\begin{align}
  \label{eq:sb}
  s_{B} = \frac{1}{\sqrt{\gamma Ra}}.
\end{align}
Figures \ref{fig:kTe100} and \ref{fig:kTe5000} show the permeability fields, temperature, the normalised determinant of the strain tensor ($|\mathbf{E}|/\overline{|\mathbf{E}|}$, where $\overline{|\mathbf{E}|}$ is the spatial average) and the modulus of the velocity $|q|$ for two selected $Ra$ and different heterogeneous cases. The homogeneous $Ra=1e2, 5e3$ cases illustrate the relation between the temperature pattern and the strain rate. The maximum strain rate is found at the stagnation points at the top and bottom boundaries. At the stagnation points, the interface is compressed, which increases the temperature gradient and the conductive heat flux forming the temperature plumes. Heterogeneous cases, however, display a more complex structure. The determinant of the strain tensor is more sensitive to the size of the heterogeneity. For small $\lambda_{x}$, the behaviour resembles that of the homogeneous cases. Fluid is highly stretched at the stagnation points where flux increases and plumes originate. There is also significant stretching along the convection rolls and plumes, although it has no effect on heat fluxes. For high $\lambda_{x}$ cases, the structure of  $|\mathbf{E}|$ no longer follows the temperature patterns but instead it mimics the permeability structure. Maximum values of $|\mathbf{E}|$ coincide with long, horizontal, high permeability areas where fluid flow changes direction and moves parallel to the $x$ direction. This causes the columnar patterns to be distorted. A closer inspection of the fluid structure near the boundary (Figure \ref{fig:strain_detail}) reveals that the formation of proto-plumes is still linked to the areas where high strain is found. The location of the high strain areas is found at high permeability regions where a rapid change in velocity at the stagnation point favors interface compression. 


\begin{figure}
    \centering    
    \includegraphics[width=\linewidth]{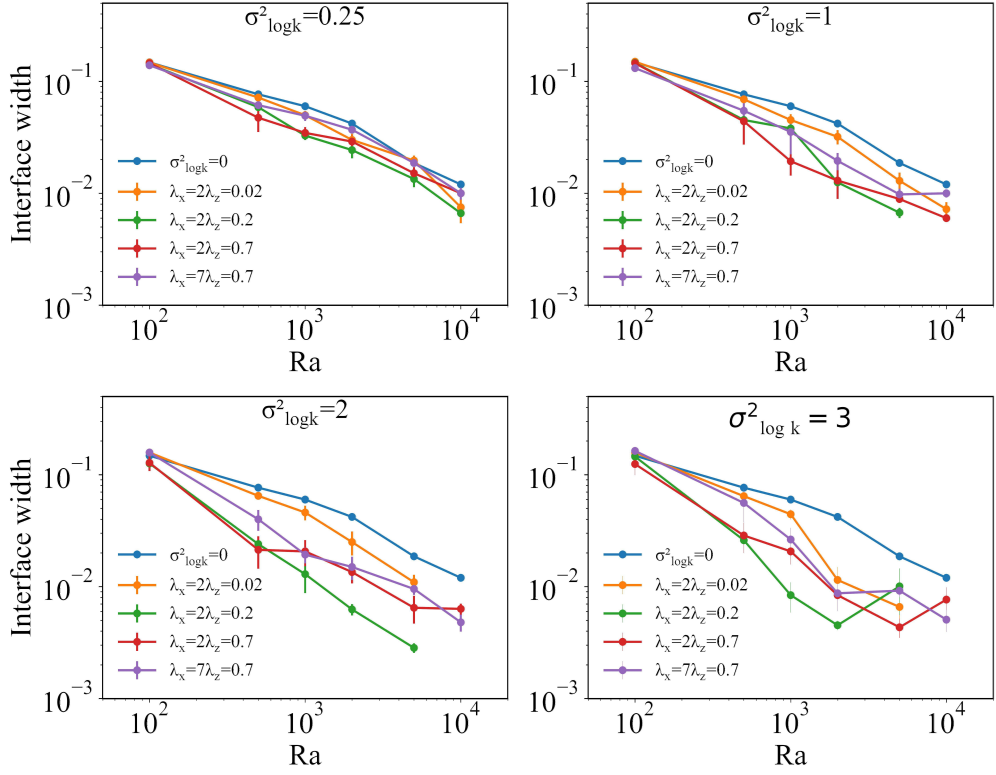}  
    \caption{Interface width versus $Ra$ for different correlation lengths and $\sigma_{\log k}^{2}$. The interface width is smaller and decreases faster with $Ra$ as $\lambda_x$ and $\sigma^2_{\log k}$ increase, meaning that the interface is more compressed in heterogeneous cases due the presence of high and low permeability regions}
   \label{fig:InterfaceSigma}
\end{figure}

To estimate the interface width $s_{B}$, we calculate the second central moment of $T(1 - T)$ at all $x$ positions along the top boundary. The interface width is then taken as the square root of the minimum calculated value to reduce the effect of the plume and stagnation points lateral movement in the high $Ra$ cases. It can be seen (Figure~\ref{fig:InterfaceSigma}) that the interface width tends to get steeper and to decrease as $Ra$ and $\sigma_{\log k}^2$ are increased and that the value it takes is smaller than in the homogeneous case suggesting that the interface undergoes greater compression in presence of a heterogeneity. As it has been previously pointed out for the fluxes, the homogeneous and the heterogeneous case with the smallest correlation length $\lambda_{x}=0.02$ behave similarly. However, the decrease in the interface width is more accentuated in the case where $\lambda_{x}=2\lambda_{z}=0.7$ and $\lambda_{x}=2\lambda_{z}=0.2$. Moreover, a higher $\sigma_{\log k}^2$ further widens the gap between the plots with different $\lambda_{x}$.

In the vicinity of a stagnation point located at a fluid interface, temperature can be modelled with an advection–diffusion equation \citep{Ranz1979, Villermaux_2012, Le_Borgne_2013}. Following \cite{hidalgo_dissolution_2015}, and using the equivalence between $\langle \chi \rangle$  and $F$ \eqref{eq:flux}, at the steady state we obtain
\begin{align}
  \label{eq:chi_stag_point}
  F = \frac{\omega_{e}}{\sqrt{4\pi}} \frac{\left(1 - T_{b}\right)^{2}}{s_{B}Ra},
\end{align}
where $\omega_{e}$ is an effective interface length and $T_{b}$ the temperature far from the interface.

For homogeneous cases, \eqref{eq:chi_stag_point} can be used to obtain the scaling of the interface width and strain rate. In conduction-dominated systems, $F \propto Ra^{-1/2}$ and $\omega_{e}$ is weakly dependent on $Ra$ because it is inversely proportional to the number of convection rolls. Thus, from \eqref{eq:sb} and \eqref{eq:chi_stag_point}, we obtain that the interface width and strain rate scale as $s_{B} \propto Ra^{-1/2}$ and $\gamma \propto Ra^{0}$. When convection dominates $F \propto Ra^{0}$. The effective interface length $\omega_{e}$ is independent of $Ra$ because it is proportional to the number of stagnation points ($\propto Ra$) and the individual length associated to each stagnation point (proportional to the wavelength of the most unstable mode, i.e., $\propto Ra^{-1}$). Therefore, in the convection-dominated regime, we obtain $s_{B} \propto Ra^{-1}$, $\gamma \propto Ra^{1}$. The change in the scaling of $\gamma$ indicates the transition from a regime in which velocity changes happen at a scale of the size of the domain (therefore $\gamma$ does not depend on $Ra$) to a regime in which velocity changes at the scale of $s_{B}$ and $\gamma$ is proportional to $Ra$, indicating that conduction and convection are coupled \citep{hidalgo_mixing_2018}.

As observed for flux, $s_{B}$ presents a transition for the homogeneous case around $Ra = 1e3$ (see figures \ref{fig:FluxRahom} and \ref{fig:InterfaceSigma}) from a conduction-dominated regime where $s_{B}$ scales between $\propto Ra^{-0.4}$ and $Ra^{-0.78}$ (similar to the  $Ra^{-0.44}$ and $Ra^{-0.98}$ scaling found by \citet{hidalgo_mixing_2018}, who simulated a larger range of $Ra$ values). The heterogeneous cases, however, do not display a distinct change in slope except for the case with the smallest correlation length $\lambda_x=2\lambda_z=0.02$ which presents a transition from  $s_{B}$ scaling between $Ra^{-0.55}$ and $Ra^{-0.4}$ (the higher the variance, the higher the absolute value of the exponent) to $s_{B}$ between $Ra^{-1.2}$ and $Ra^{-0.78}$. The rest of the heterogeneous cases do not display a distinct change in slope. The fitted exponents for $s_{B}$ range between -0.974 and -0.55 ($\lambda_{x}=2\lambda_{z}=0.2$ between -0.974 to -0.65;  $\lambda_{x}=7\lambda_{z}=0.7$ between -0.78 and -0.55; and  $\lambda_{x}=2\lambda_{z}=0.7$ between -0.7 and -0.56). The scaling of $s_{B}$ is more difficult to interpret. The variety of scaling suggests a combined effect of the heterogeneity on the velocity structure. On one hand, the behaviour of $s_{B}$ in the heterogeneous cases indicates that, even for high $Ra$, velocity changes do not happen at the scale of $s_{B}$ but at an intermediate scale influenced by the permeability structure. On the other hand, the number of stagnation points and their associated length depends on $Ra$ and the correlation lengths of the permeability field. Stagnation points do not distribute evenly and their location depends on the local permeability structure (see, for example Figure \ref{fig:kTe5000}c and e). This coincides with the observations on the instability modes by \citet{Mckibbin1982} and \citet{Rees2009}.
%
%
\subsection{Mixing state}
The mixing state, which shows how the temperature is redistributed within the system, is given by the probability density function (pdf) of the temperature
\begin{align} 
\label{eq:pdf}
    p(T)=\frac{1}{A}\int_\Omega \delta[T - T(\mathbf{x})] d\Omega
\end{align} 
and the segregation intensity \citep{Danckwerts1952}
\begin{align} 
\label{eq:intensseg}
    I = \frac{\sigma_{T}^{2}}{\overline{T}(1-\overline{T})},
\end{align} 
where $A$ is the area of the domain, $\delta(T)$ denotes the Dirac delta, $\overline{T}$ is the mean of the temperature and $\sigma^{2}_{T}$ its variance. The intensity of segregation ranges between 0 and 1 and indicates the homogeneity of a mixture. When $I=1$, the system is completely segregated and when $I=0$ the temperature distribution is uniform. In the following, $p(T)$ and $I$ are time averages within the convection-dominated regime of the mean of five realisations. 
\begin{figure}
    \centering   
    \includegraphics[width=\linewidth]{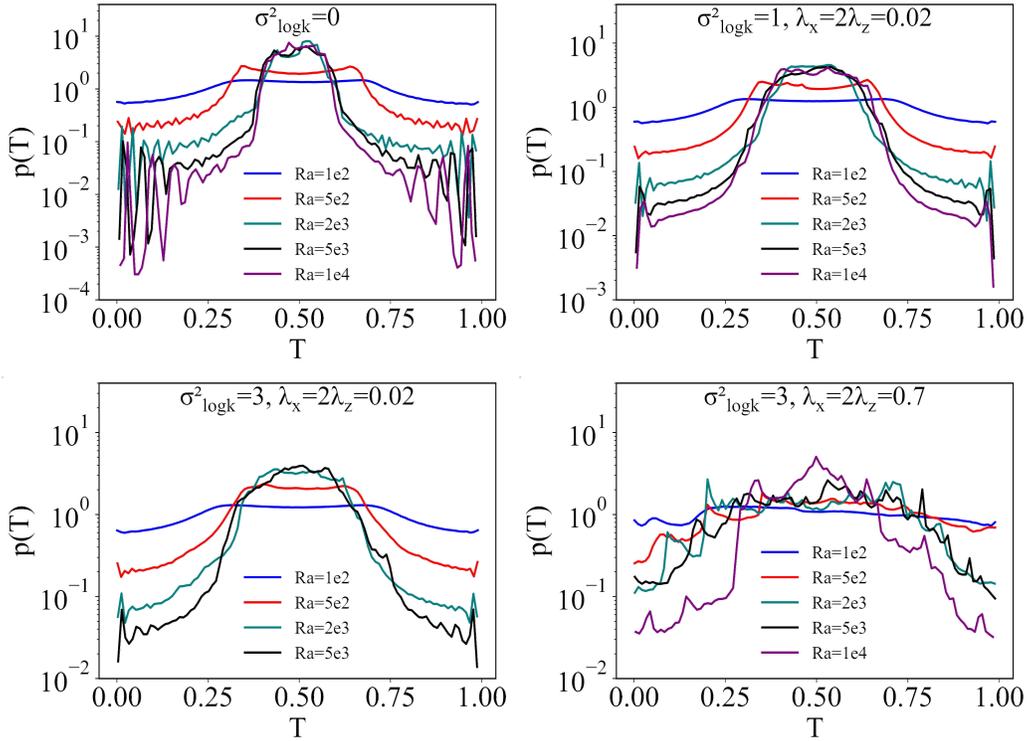}  
   \caption{Temperature probability density functions with different $\sigma_{\log k}^{2}$. Increasing Ra, narrows the pdfs whereas increasing $\sigma_{\log k}^{2}$ flattens their peak thus making them wider particularly with higher $\lambda_x$}
   \label{fig:PDFs}
\end{figure}

\begin{figure}
    \centering    
    \includegraphics[width=\linewidth]{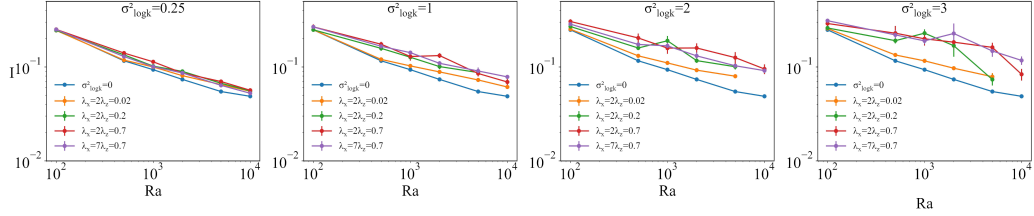}   
    \caption{Time-averaged intensity of segregation during the convection-dominated regime ($60 <t< 250$) for different $\lambda_{x}$ and $\sigma_{\log k}^{2}$. Increasing $Ra$ decreases $I$. All considered heterogeneous cases are more segregated than the homogeneous one. The intensity of segregation gets bigger as $\lambda_{x}$ and $\sigma_{\log k}^{2}$ increase. Strongly heterogeneous cases (high $\sigma_{\log k}^{2}$) are hence more segregated than weakly heterogeneous ones  (low $\sigma_{\log k}^{2}$)}
    \label{fig:SegregationIndex}
\end{figure}

The temperature probability density functions and the segregation index (figures~\ref{fig:PDFs} and \ref{fig:SegregationIndex}) reflect the effect of the pattern formation in the mixing state of the system. For low $Ra$ and weak heterogeneity, the system is organised in convection rolls, which maintain high and low $T$ areas separated. this is reflected in wide temperature pdfs wide high $I$. As $Ra$ increases and the temperature patterns become distorted and more chaotic, convection favors fluid mixing, the pdfs narrow and the segregation index diminishes. Columnar patterns lead to less fluid segregation than convection rolls. Heterogeneity however, alters the temperature patterns creating preferential paths and, in some cases, reverting the fluid structure from columnar patterns to convection rolls. Accordingly, fluid segregation increases $I$ compared to the homogeneous case, and the temperature pdfs become broader with flatter peaks that their homogeneous counterparts. This effect is particularly pronounced at high values of $\lambda_{x}$, which induce a barrier effect that increases the interfacial area and thereby hinders mixing. 
Furthermore, increasing $\sigma_{\log k}^{2}$ makes the heterogeneous cases more segregated by increasing $I$ and flattening the peak of the pdfs.
%
%
\section{Summary and conclusions}
Convective mixing in heterogeneous porous media has been investigated in the context of the Horton-Rogers-Lapwood problem, where convection is triggered by a Rayleigh-Bénard instability. A parametric study was conducted by varying the Rayleigh number ($Ra$), the variance of the log-normally distributed permeability field ($\sigma^2_{\log k}$),  and its correlation length in the longitudinal direction ($\lambda_{x}$). The results indicate that increasing $\lambda_{x}$ and $\sigma^2_{\log k}$  leads to more dispersive fingering patterns and enhances the heat flux. Likewise, reducing the anisotropy ratio results in higher flux. The dependency of flux on $Ra$ changes from conduction-dominated to convection-dominated in weakly heterogeneous and homogeneous ones. The strongly heterogeneous cases do not show a change of regime for the simulated $Ra$ cases.

The relation between temperature patterns and velocity structure is also affected by the heterogeneity. The location of the formation of temperature plumes coincides with stagnation points where the interface is compressed. In homogeneous and weakly heterogeneous cases, the location of the stagnation points is related to the shape of the convection rolls and temperature plumes. In strongly heterogeneous media, these stagnation points are linked with the heterogeneity structure and they are found near high permeability areas. At this stagnation points, the interface width decreases as $Ra$, $\sigma^2_{\log k}$ and $\lambda_{x}$ increase, with the fluid interface undergoing greater stretching in high-permeability regions. Fingering patterns tend to align with high-permeability pathways, causing the interface to deform more significantly than in homogeneous cases. 

The segregation intensity and the variance of the temperature probability density functions, increase with larger $\lambda_{x}$ and $\sigma^2_{\log k}$, taking significantly higher values than those observed in homogeneous cases. This suggests that greater heterogeneity results in stronger segregation effects and less well-mixed systems.

These findings highlight the role of permeability heterogeneity in governing convective mixing in porous media and provide scalings with $Ra$ and the heterogeneity parameters that can be used to estimate heat fluxes in geothermal applications. The interplay between $\lambda_x$, $\sigma^2_{\log k}$ and $Ra$ not only controls the fingers morphology but also dictates the mixing state and efficiency. It is thus paramount to account for structural heterogeneity in predictive models of natural and engineered systems where accurate flux estimation and interface stability are essential.
%
%
\paragraph{\textbf{Acknowledgments}}
RB and JJH acknowledge the H2020 MSCA ITN program under the Grant No. 956457 (COPERMIX) and the support of the MICIU/AEI/10.13039/501100011033 and the European Union NextGenerationEU/PRTR through grant CNS2023-144134 (ESFERA).
%
%
\appendix
\section{Numerical setup and convergence analysis}\label{app:numerics}
The governing equations were solved using the open-source CFD toolbox OpenFOAM \citep{Weller1998} using a second-order scheme backward for time derivatives, Gauss linear for gradients, and Gauss linear orthogonal for Laplacians. The resulting linear systems of equations were solved with using GAMG solver for the pressure equation and a PBiCGStab for the temperature equation with an absolute tolerance of $10^{-10}$ and a relative tolerance of $10^{-3}$. We prescribed a tight residual control $(10^{-8})$ for the PIMPLE algorithm used for the pressure-velocity-temperature coupling.
    
Simulations were performed on a high-quality, structured, hexahedral, uniform mesh with no grading, which ensures high orthogonality and avoids numerical artifacts associated with distorted cells.

Mesh convergence was performed by comparing the mean steady-state heat flux at the convection-dominated regime for three successively refined meshes, each with an aspect ratio of $1.5$. The total number of cells ranged from $1.8 \times 10^{5}$ (coarse) to $9.1 \times 10^{5}$ (fine), corresponding to grid resolutions of $600 \times 300$, $900 \times 450$, and $1350 \times 675$. The greatest relative change in fluxes between two successive refinement levels was less than a few percent. We considered this difference a good compromise between numerical accuracy and computational cost.

We also ensured that the meshes for the different cases were fine enough to resolve the most unstable wavelength ($\lambda$) as predicted by linear stability analysis. This $\lambda_{c}$ is considered the smallest length scale of the instability. The critical wavelength is given by $\lambda_{c} = 90L_{z}/Ra$ \citep{riaz_2006, Hidalgo2013}. All used meshes allocated more than two cells per wavelength to capture the instability.

For the heterogeneous cases, we also ensured that our meshes were sufficiently fine to resolve the fine-scale details of the heterogeneity, with a minimum of five cells allocated per correlation length. The permeability fields were generated with the utility \texttt{setRandomField} of the SECUReFOAM platform. This utility uses a stochastic Fourier integral to represent the Gaussian random field as a sine and cosine series with coefficients depending on the semivariogram.
%
%
%

%
%
\end{document}